\documentclass{ws-procs10x7}

\def\PRL{Phys. Rev. Lett. }
\def\PRC{Phys. Rev. C }
\def\PRD{Phys. Rev. D }
\def\PLB{Phys. Lett. B }
\def\NPA{Nucl. Phys. A }
\def\ptt{p$_{\rm t}$\,}
\begin{document}

\title{Has the Quark-Gluon Plasma been seen?}

\author{J. Stachel}

\address{Physikalisches Institut, Ruprecht-Karls-Universit{\"a}t Heidelberg,
  Philosophenweg 12,\\ D 69120 Heidelberg, Germany\\E-mail: stachel@physi.uni-heidelberg.de}

\twocolumn[\maketitle 
]

\section{Introduction}

Very shortly after the discovery of asymptotic freedom~\cite{asfree} it became apparent
that, as a consequence, at high temperature and/or at high density quarks and
gluons would also become deconfined~\cite{asqcd}, leading to a phase transition from
confined hadronic matter to an unconfined phase. This was studied in
subsequent years and since the early 1980ies this phase is called the
Quark-Gluon Plasma (QGP). 

The conditions for this phase transition were studied in lattice QCD and state
of the art calculations~\cite{karsch} obtain as critical temperature for the phase
transition for two light and one heavier quark flavors a value for the
critical temperature of T$_c$ = 173 $\pm$ 15 MeV and for the critical energy
density of $\epsilon_c$ = 0.7$\pm$ 0.2 GeV/fm$^3$. It is believed since many years that
in collisions of heavy atomic nuclei at high energies such conditions should
be reached. This motivated an experimental program starting simultaneously in
1986 at the Brookhaven AGS and at the CERN SPS, initially with light
projectile nuclei such as Si and S and from 1992 and 1994, respectively, with
Au and Pb projectiles. The experimental results from this program prompted a
press release from CERN~\cite{cernpress} in February 2000 stating that the combined
results from the experiments proved that a new state of matter other than
ordinary hadronic matter had been created in these collisions, in which
quarks were 'liberated to roam freely'. The experimental results were clearly
not reconcilable with the known hadronic physics and it could be estimated that
the critical temperature had been exceeded in the early phase of the collision
by about 20-30 \% and the critical energy density by somewhat more than a
factor 2. On the other hand, from those data nothing could be said yet that
would characterize the properties of the new state of matter. Hence, at that
time the term QGP was not used for the new state of matter.

In the summer of 2000, RHIC as a dedicated collider for heavy ions started
operation with two large experiments, PHENIX and STAR, and two smaller
experiments, BRAHMS and PHOBOS. In the first 3 years of operation data for Au
+ Au collisions with an integrated luminosity of 85/$\mu$b, for p + p
collisions with 2/pb, and for d + Au collisions with 25/nb were collected and
a summary of the results was recently published in a special issue of Nuclear
Physics A by all four heavy ion
experiments~\cite{brahms,phobos,star,phenix}. In the 2004 Au + Au run the 1/nb
level was exceeded and data start to appear from this run. Here I will rely
mostly on published data and review some of the key observations from the
first 3 years including only a few of the still preliminary first run4
observations.

\section{Experimental Results}

\begin{figure*}
\epsfxsize35pc
\figurebox{16pc}{32pc}{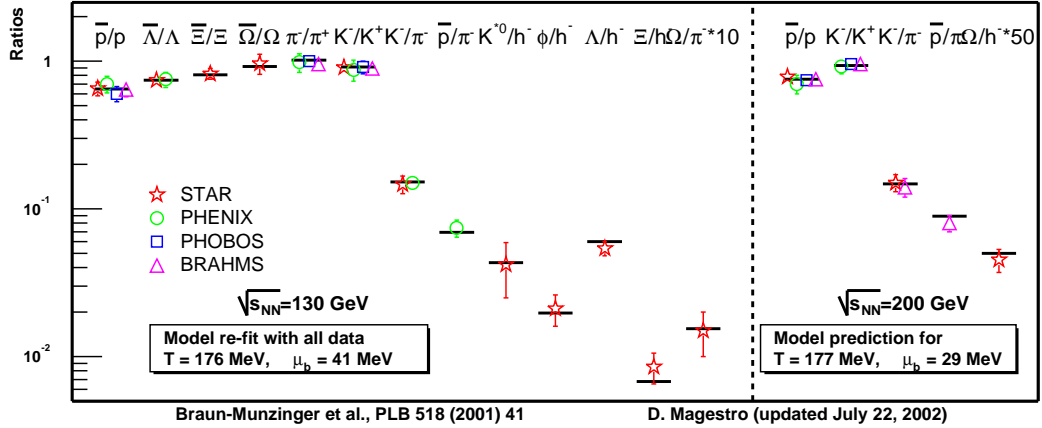}
\caption[]{Hadron yield ratios measured at RHIC in comparison to calculations
  within a statistical model based on a grand canonical ensemble (updated
  version of~\cite{rhicthermal}, taken from~\cite{hwareview}).} 
\label{fig:yields}
\end{figure*}

\subsection{Hadron Production and Statistical Models}

Hadron yields have been measured for a large range of species at the AGS, SPS
and at RHIC. It was realized already for many years that the data for central
collisions of heavy nuclei can be rather accurately reproduced by calculations
for a chemically equilibrated system in terms of a grand canonical ensemble (a
review and complete set of references can be found in~\cite{hwareview}). For
the lower RHIC energy of $\sqrt{\rm s}$ = 130 GeV the data are final and
published and for 200 GeV data are emerging currently. Figure~\ref{fig:yields}
shows experimental yield ratios from all four RHIC experiments in comparison
to a statistical model fit.

\begin{figure}
\epsfxsize220pt
\figurebox{120pt}{160pt}{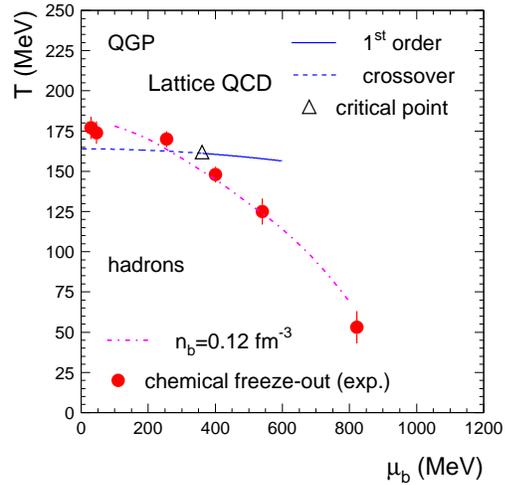}
\caption[]{Phase diagram of nuclear matter in the temperature - baryon chemical
  potential plane. Experimental points for hadro-chemical freeze-out are shown
  together with a recent lattice QCD calculation~\cite{lattice2} and a curve
  of constant total baryon density. Figure from~\cite{phasedia}.}
\label{fig:phase}
\end{figure}

 In the calculations, there are two free fit
parameters, the temperature and the baryon chemical potential. For top RHIC
energy the temperature is fitted as 177$\pm$ 5 MeV, practically unchanged from
$\sqrt{\rm s}$ = 130 and 17.3 GeV; the baryo-chemical potential is dropping
continuously with increasing beam energy reflecting an increasing transparency
of the nuclei at higher energies and an increasing dominance of
baryon-antibaryon production. This is shown in Figure~\ref{fig:phase} where
results of statistical model fits at various beam energies are summarized and
shown together with recent results from lattice QCD~\cite{lattice2}.

\begin{figure*}
\epsfxsize34pc
\figurebox{16pc}{32pc}{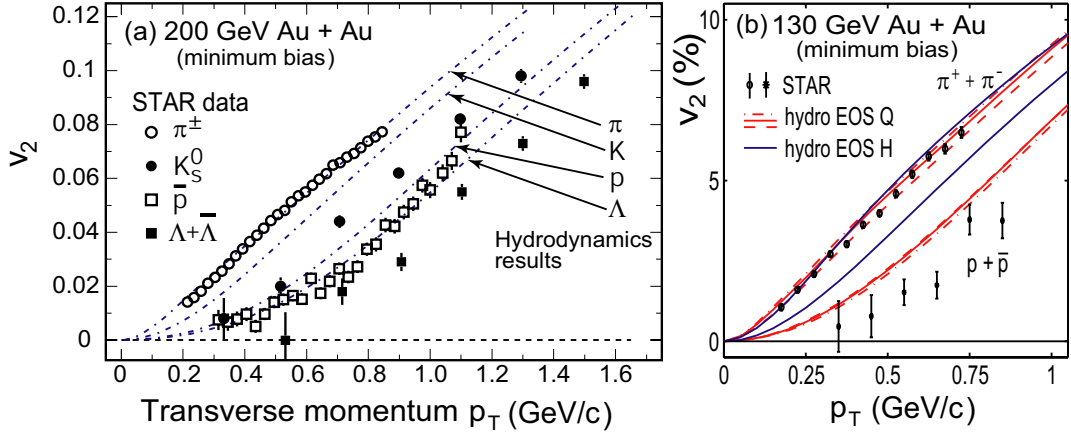}
\caption[]{Left: Elliptic flow coefficient v$_2$ as function of \ptt for
  different particle species~\cite{star200v2}. Together with the experimental
  data results from a hydrodynamics calculation including a phase transition
  are shown~\cite{hydrov2}. Right: Experimental data for pions and protons at
  a lower RHIC energy~\cite{star130v2}. Also shown are hydrodynamics
  calculations~\cite{hydrov2} with and without phase transition from QGP to
  hadronic phase. Figure from~\cite{star}.}
\label{fig:v2}
\end{figure*}

It appears that from top SPS energy upwards the temperature at which
hadro-chemical equilibrium is achieved is not changing anymore and practically
coincides with the lattice QCD prediction for the critical temperature, while
at lower beam energies it is falling. At $\sqrt{s}$ = 8.8 GeV it is only 148
$\pm$ 5 MeV. The strangeness suppression that is well established for pp and
e$^+$e$^-$ collisions appears to be completely lifted. This leads to an
enhancement in the yields of particularly multistrange hadrons in heavy ion
collisions as compared to pp results. For the Omega baryon at SPS energy this
enhancement is~\cite{wa97omega} a factor 17. How hadrons like the $\Omega$ can
be equilibrated on the time scales of the nuclear collision has been a puzzle
for several years and there is consensus that with two-body collisions and the
known hadronic cross sections this is not
possible~\cite{rapidequ,greiner,kapusta}. A possible explanation has been
presented recently~\cite{rapidequ}. In the direct vicinity of the phase
transition the densities of particles are rising very rapidly due to the
increase of degrees of freedom by more than a factor of 3 between a hadron gas
and a QGP. At these high densities multi-hadron collisions become dominant and
can drive even the $\Omega$ yield into equilibrium in a fraction of a fm/c.
Conversely, already 5 MeV below the critical temperature the densities are so low
that the system falls out of equilibrium and the yields cannot follow anymore
a decreasing temperature. Therefore the authors of~\cite{rapidequ} conclude
that the rapid equilibration is a direct consequence of the phase transition
from QGP to hadronic matter and that, at least at high beam energies, the
chemical equilibration temperature is a direct experimental measure of the
critical temperature.
 
\subsection{Elliptic Flow}

\begin{figure*}
\epsfxsize200pt
\figurebox{16pc}{32pc}{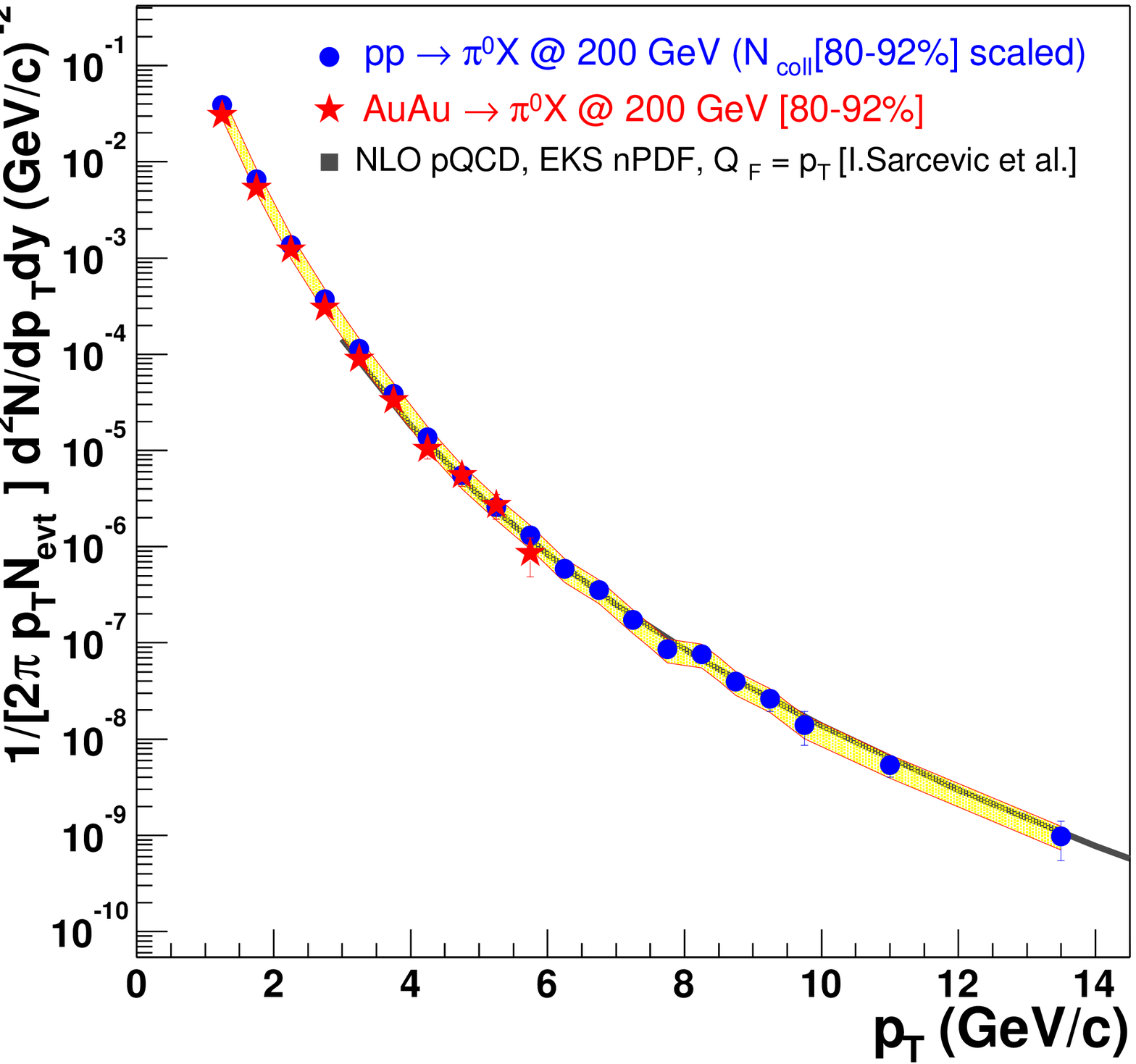}
\epsfxsize200pt
\figurebox{16pc}{32pc}{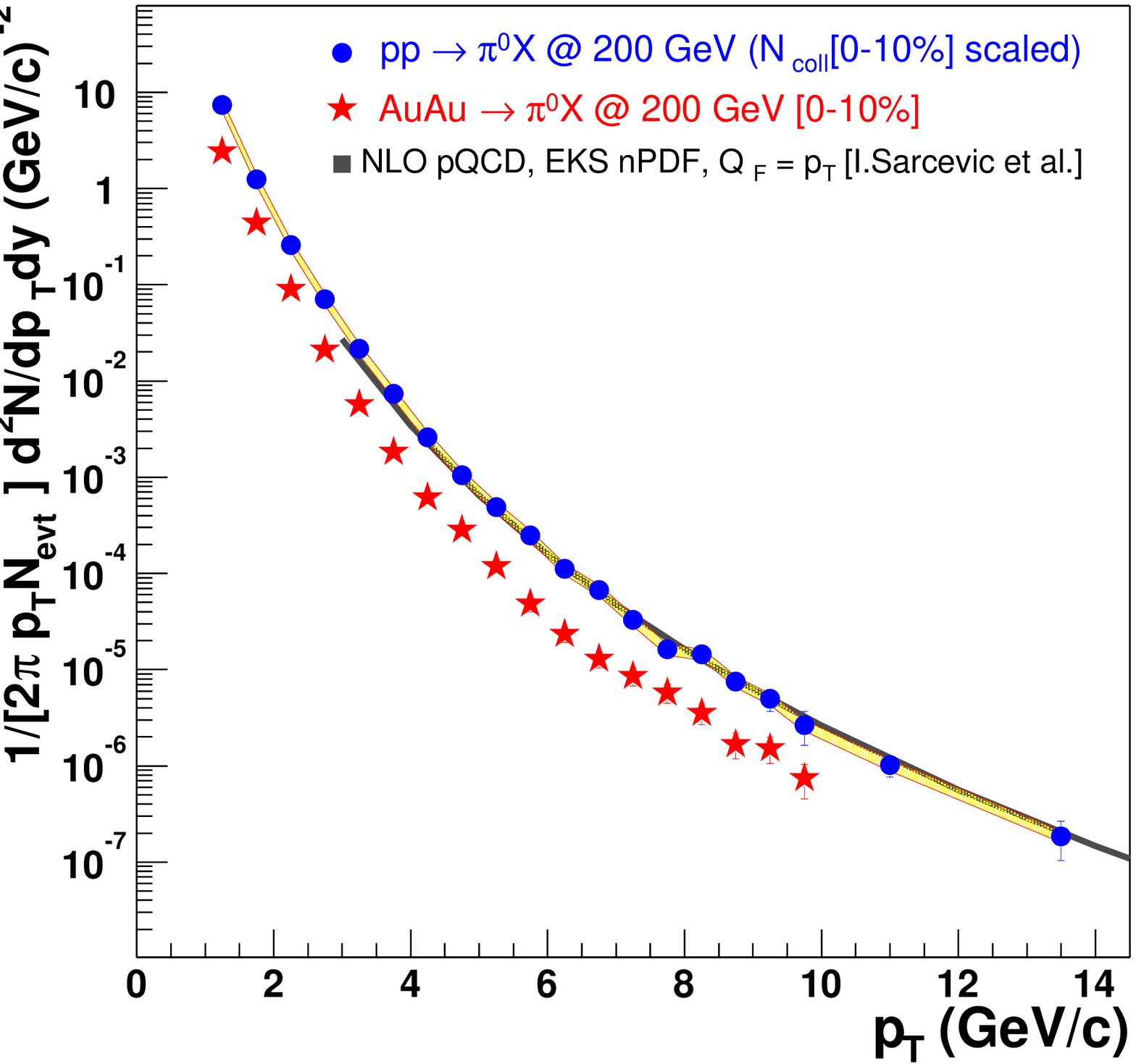}
\caption[]{Neutral pion transverse momentum spectra measured by PHENIX in peripheral
  (left) and central (right) Au + Au collisions (stars) together with pp data
  from the same experiment scaled with the number of binary collisions
  (circles)~\cite{phenixhipt,phenixhiptpp}. Yellow band: Normalization
  uncertainties of the pp data. Black line: NLO pQCD
  calculation. Figure taken from~\cite{denterria}. }
\label{fig:spectra}
\end{figure*}

Momentum distributions in three dimensions are analyzed with transverse
coordinates relative to the reaction plane of the collision spanned by the
impact parameter vector and the beam direction and a decomposition in terms of
Fourrier coefficients is performed. Already at the Bevalac sizeable
anisotropies were observed for heavy colliding nuclei. In particular, the
quadrupole coefficient v$_2$ was found to be negative, explained by shadowing
of the emitted particles by the target and projectile spectator
remnants~\cite{plasticball}. At AGS energies a sign change was
observed~\cite{flowe877} by E877, i.e. the momentum spectra were harder in the
reaction plane than perpendicular to it. The interpretation used a prediction
from hydrodynamics~\cite{hydro} that, for semiperipheral collisions, in the
early phase of the collision the pressure gradient was larger in this
direction due to the excentricity of the nuclear overlap region\footnote{The
  use of hydrodynamics to describe the dynamics of a hadronic collision goes
  back to the 1950ies~\cite{landau}.}. From the hydrodynamic evolution it
would follow, that this anisotropy in pressure gradient would evolve with time
into an anisotropy in momentum space, driven by the initial condition and the
equation of state of the expanding system. This was confirmed by a microscopic
analysis within a transport model~\cite{sorge}. From this the name 'elliptic
flow' originated for the quadrupole coefficient v$_2$.

At the higher SPS energy growing positive coefficients v$_2$ were
found~\cite{v2na45,v2na49,v2wa98} and the sign change was traced to
occur~\cite{v2e895} at beam momenta per nucleon of about 4 GeV/c. At RHIC
energies very large values of v$_2$ were
observed~\cite{star130v2,star200v2,phenixv2,phobosv2}, typically about 50~\%
above SPS  top energy results.

This was studied differentially for different hadronic species and as function
of p$_{\rm t}$ as shown for data from STAR in Figure~\ref{fig:v2}. It is
observed that for more massive hadrons the rise of v$_2$ starts at larger
values of p$_{\rm t}$. For the first time there was quantitative agreement
with hydrodynamic calculations~\cite{hydrov2,v2teaney} in terms of p$_{\rm t}$
and hadronic species dependence, as also shown in Fig.~\ref{fig:v2}. These
hydrodynamic calculations also reproduce the overall features of the \ptt
spectra of differenct hadrons, although in details there are deviations
stemming from the different treatment of the hadronic phase and freeze-out
(see Fig. 20 in~\cite{phenix} and references there). It is common to all the
hydrodynamics calculations that, in order to reproduce the data, a rapid
initial equilibration on a time scale faster than 1 fm/c is
required~\cite{hydrov2,v2teaney,hydroreview}. 

At \ptt above 2-3 GeV/c, where hydrodynamics should no longer hold as a
theoretical description, another type of scaling was
discovered~\cite{v2cv,star}: dividing both v$_2$ and \ptt by the number of
constituent quarks in a hadron all results match rather well even including
multistrange baryons. It was realized that an old idea of quark
coalescence~\cite{hwa} could be the underlying physics~\cite{qcoal} and indeed
calculations based on the assumption of coalescence of valence quarks during
hadronization of a QGP reproduce this feature rather well~\cite{qcoal2}.

\subsection{High Momentum Suppression}

One of the highlights of the RHIC experimental program is the observation of a
strong suppression in the production of hadrons at high transverse momentum
when compared to pp collisions. Figure~\ref{fig:spectra} shows the \ptt
spectrum of neutral pions in Au + Au collisions as compared to a measurement
in pp in the same experiment and at the same
energy~\cite{phenixhipt,phenixhiptpp}. The pp spectrum compares well with a
calculation in NLO pQCD. In order to compare, the pp spectrum has been scaled
with the number of binary nucleon-nucleon collisions in a Au + Au collisions
at a given centrality. The number N$_{\rm coll}$ of binary collisions is given
by the collision geometry - measured in the data with some resolution -, the
well known nuclear density distribution, and the inelastic pp cross section.
The collision centrality in Au + Au collisions is characterized by the
fraction of the geometric cross section for which events have been selected,
which is related to the impact parameter. The yellow bands in
Fig.~\ref{fig:spectra} reflect the systematic uncertainty in this scaling. One
can observe that for peripheral collisions pp and Au + Au collisions agree
very well, while in central collisions the Au + Au spectrum is significantly
suppressed.

\begin{figure}
\epsfxsize190pt
\figurebox{120pt}{160pt}{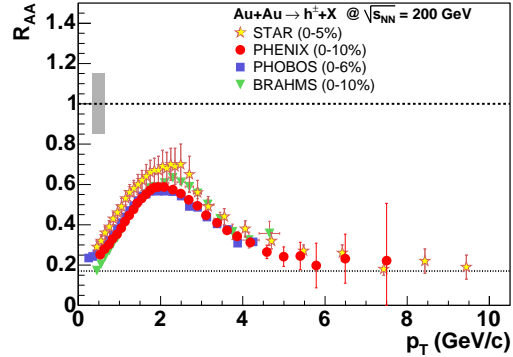}
\caption[]{Ratio R$_{\rm AA}$ of the \ptt spectrum for central Au + Au collisions
  normalized to the pp spectrum scaled with the number of binary collisions
  for charged particles from all four RHIC experiments. Figure
  from~\cite{denterria2}.}
\label{fig:raa}
\end{figure}

This is better visualized by building the ratio R$_{\rm AA}$ between the Au +
Au \ptt spectrum and the pp spectrum scaled with N$_{\rm coll}$ as shown in
Figure~\ref{fig:raa}. All four RHIC experiments observe a suppression by about
a factor of five for \ptt larger than 4 GeV/c. Since not all experiments
measure neutral pions, the ratio is shown here for charged hadrons, but at
large \ptt the data for all hadron species merge. At low \ptt the ratio
R$_{\rm AA}$ is expected to be below one because there, due to the dominance of
soft processes, the appropriate scaling is with the number of participants,
i.e. nucleons in the nuclear overlap region. It is expected that this ratio
should rise as hard scattering becomes dominant and, in fact, due to the well
known Cronin enhancement, in the region of 2-6 GeV/c values above one are
expected. Contrary to this expectation the data show a suppression.

\begin{figure}
\epsfxsize200pt
\figurebox{120pt}{160pt}{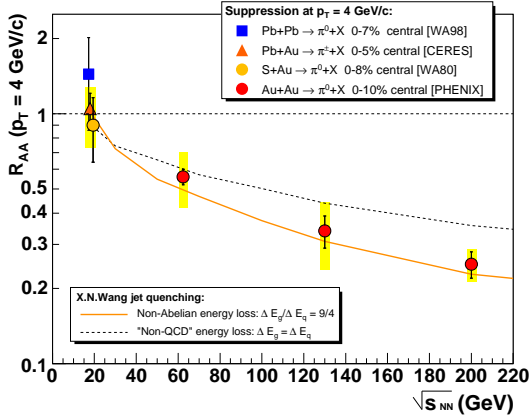}
\caption[]{Suppression factor R$_{\rm AA}$ for pions as function of beam energy
at a fixed \ptt value of 4 GeV/c. The yellow line shows a
calculation~\cite{wangwang} including parton energy loss in a medium with high
parton density. Figure from~\cite{denterria3}. }
\label{fig:raaroots}
\end{figure}

The suppression is not unexpected. It was predicted that in a medium with high
parton density the radiative energy loss of a quark or gluon should be
strongly enhanced~\cite{wang,baier}, leading to a very effective
thermalization of jets in a hot color charged medium. Calculations employing a
large initial gluon rapidity density of about 1100 can account~\cite{vitev}
for the data at top RHIC energy. The beam energy dependence of the
R${\rm_{AA}}$ ratio was presented recently by d'Enterria~\cite{denterria3}
and it appears that the suppression evolves in a very smooth way from top SPS
energy onwards. The R${\rm_{AA}(p_t=4 GeV/c)}$ values are shown in
Figure~\ref{fig:raaroots} for the top SPS energy and three RHIC
energies. Already the values of about 1.0 measured at the SPS represent a
slight suppression as compared to the normal Cronin
enhancement~\cite{denterria3}. Going from $\sqrt{\rm s_{nn}}$ = 17.3 to 62.4
to 200 GeV the gluon rapidity density needed to reproduce the data
grows~\cite{vitev} from 400 to 650 to 1100. An alternative formulation of this
in medium suppression is by increasing and large opacities of the medium
traversed~\cite{wangwang}.

The proof that this is really a final state effect probing the properties of
the medium traversed by the parton is given by the observation that in d + Au
collisions in the same experiments no suppression is seen, but rather the
expected Cronin enhancement~\cite{dauphobos,dauphenix,daustar,daubrahms}.

\begin{figure}
\epsfxsize200pt
\figurebox{120pt}{160pt}{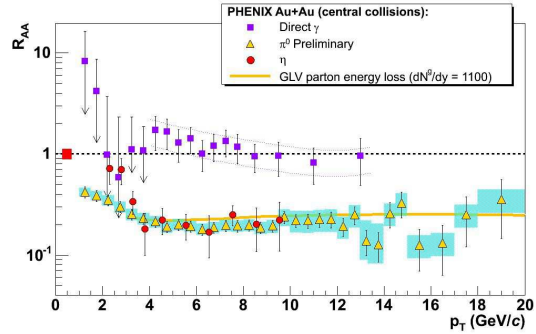}
\caption[]{Preliminary PHENIX results for the suppression factor R$_{\rm AA}$
out to high \ptt for $\pi^0$ and $\eta$ mesons together with a calculation
based on a high gluon rapidity density~\cite{vitev}. Also shown are the
results for direct photons. Figure from~\cite{akiba}.}
\label{fig:raahighpt}
\end{figure}

Direct photons were measured by PHENIX in pp and Au + Au collisions at top
RHIC energy~\cite{phenixgammapp,phenixgammaauau}. The pp spectra are rather
close to a NLO pQCD calculation~\cite{vogelsang2}. The Au + Au photon spectra
are within errors consistent with the scaled pp result and hence the
expectation from NLO pQCD. For all centralities they do not show any
significant suppression. This is shown in Figure~\ref{fig:raahighpt} where
also the most recent neutral pion results from run4~\cite{akiba} extending out
to \ptt = 20 GeV/c are displayed. It is remarkable that the suppression of the
pion \ptt spectrum remains practically constant over a large range in \ptt
from 4 to 20 GeV/c, close to the predicted behavior for a medium with initial
gluon rapidity density of 1100 (see Fig.~\ref{fig:raahighpt}).

The high initial gluon densities correspond to an initial temperature of
about twice the critical temperature and to initial energy densities
$\epsilon_0$ = 14 - 20 GeV/fm$^3$ well in line with the initial conditions
needed for the hydrodynamics calculations to describe spectra and elliptic
flow (see previous section) and bracketed by the estimates based on the
Bjorken formula and transverse energy production. 

The observed high \ptt suppression pattern is different for different hadronic
species~\cite{brahms,phenix,star}. In particular, a pattern appears where at
intermediate values of \ptt of 2-6 GeV the suppression of baryons is
significantly weaker than that of mesons. The proton/pion or also the
$\Lambda$/K$^0_s$ ratios peak at values 1.5-1.6 for \ptt = 3-4 GeV/c, close to
the ratio 3/2 expected in quark coalescence models. 

\begin{figure}
\epsfxsize180pt
\figurebox{120pt}{160pt}{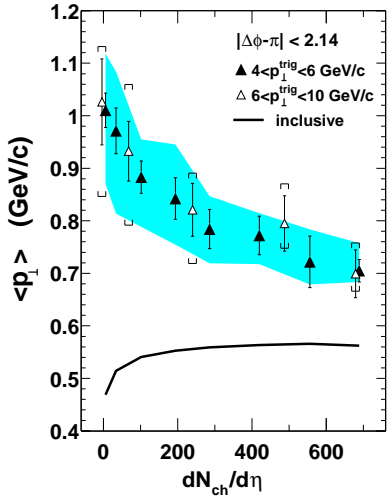}
\caption[]{Mean transverse momentum in a cone of an opening angle of one radian opposite
  to a leading particle as a function of the collision centrality;
  distributions are for two ranges of leading particle transverse momentum as
  well as for the inclusive particle distribution. Data and figure from
  STAR~\cite{startherm}.}
\label{fig:ptcone}
\end{figure}

Parton thermalization is displayed in a very clean way be recent results of
the STAR collaboration~\cite{startherm}. Evaluating the mean
transverse momentum in a cone opposite to a high \ptt trigger particle as a
funtion of centrality, a gradual decrease for more central Au + Au collisions
is observed and in the most central collisions a value very close to the
inclusive mean \ptt is reached (see Figure~\ref{fig:ptcone}).

\begin{figure}
\epsfxsize180pt
\figurebox{120pt}{160pt}{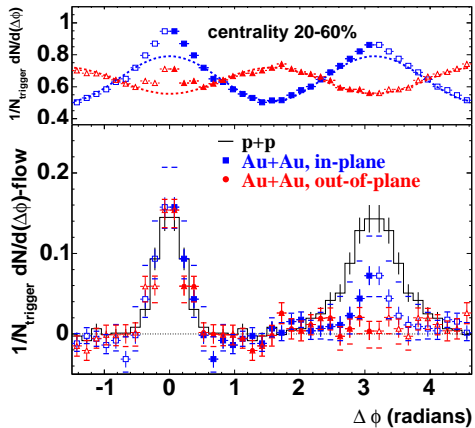}
\caption[]{Azimuthal angle correlation between a high \ptt trigger particle (4-6 GeV/c)
  with all particles in the window \ptt = 2-4 GeV/c for intermediate
  centrality Au + Au collisions. The distribution is shown for both particles
  in a $\pm$ 45 degree window around the reaction plane orientation (in-plane) or a same
  window perpendicular to it (out-of-plane). Figure from~\cite{starjet}. }
\label{fig:deltaphi}
\end{figure}

In azimuthal correlations of two high \ptt particles it was seen that the
away-side peak disappears in central Au + Au collisions for a choice of
trigger \ptt of 4-6 GeV and \ptt of the correlated particle of 2-4
GeV/c~\cite{starazim}. In pp, d + Au and peripheral Au + Au collisions a clear
peak opposite to the trigger particle is observed in the same type of
correlation, also measured by STAR~\cite{starazim,daustar}. Recently, it was
shown that the effect is very strong in case the away-side jet is emitted out
of the reaction plane and much weaker for emission in the reaction
plane~\cite{starjet} as displayed in Figure~\ref{fig:deltaphi}. This supports
the strong correlation of the suppression with the length of matter traversed
by the parton.

\begin{figure}
\epsfxsize205pt
\figurebox{120pt}{160pt}{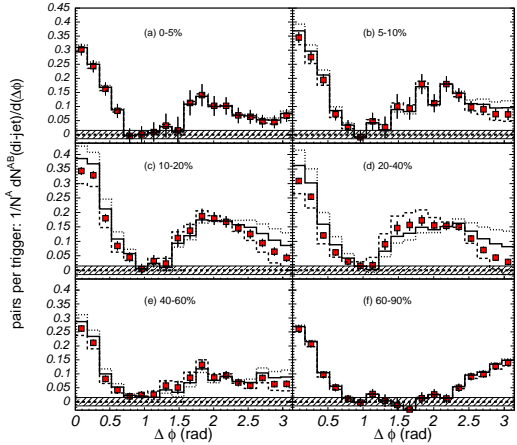}
\caption[]{Azimuthal correlations of a leading particle of \ptt = 4-6 GeV/c and any
  particle with \ptt = 1-2.5 GeV/c for different centralities of a Au + Au
  collision. Data and figure from PHENIX~\cite{phenixhole}.}
\label{fig:jethole}
\end{figure}

When lowering the \ptt cut on the correlated hadron, a very broad structure
appears on the side opposite to the trigger particle. This was shown by
STAR~\cite{startherm} for a cut on the correlated hadron of \ptt = 0.15 - 4
GeV/c. This calls to mind a similar observation at SPS energy by
CERES~\cite{ceres} where for a condition \ptt $\geq$ 1.2 GeV/c for both
particles also very strong broadening of the away-side structure with
increasing collision centrality in Pb + Au collisions was observed. Recent
data~\cite{phenixhole} from PHENIX display a tantalizing feature as shown in
Figure~\ref{fig:jethole}: For a trigger particle \ptt of 4-6 GeV/c and a
correlated particle \ptt of 1.0 - 2.5 GeV/c the away-side peak seen in
peripheral Au + Au collisions develops actually into a hole at $\Delta \phi =
\pi$ for more central collisions while a very broad peak appears with a
maximum at $\Delta \phi = \pi-1$ as can be seen in Fig.~\ref{fig:jethole}. A
suggestion has been made that this could be the Mach cone due to the sonic
boom of the quenched jet. A parton traversing a quark-gluon plasma with
velocity larger than the velocity of sound in the QGP ($\sqrt{1/3}$ for an
ideal gas) would radiate only up to a cone angle of about 1
rad~\cite{stoecker,casalderrey}. If this could be established it would have
far reaching consequences since it would be an observable linked directly to
the speed of sounds of the quark-gluon plasma and thereby its equation of
state. It remains an experimental challenge to establish an actual
cone topology in two dimensions.

\subsection{Charm Quarks and Quarkonia}

\begin{figure}
\epsfxsize195pt
\figurebox{120pt}{160pt}{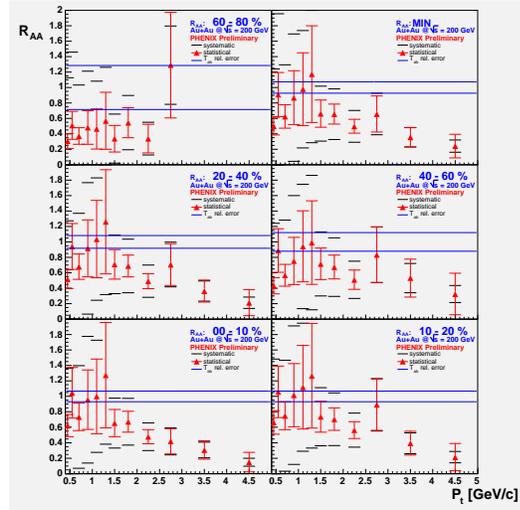}
\caption[]{R$_{\rm AA}$ suppression factor for electrons dominantly from open charm and
  open beauty decay for Au + Au collisions at top RHIC energy for different
  collision centrality (see text) measured by PHENIX. Figure from~\cite{jacak}. }
\label{fig:raacharm}
\end{figure}

Open charm has been measured indirectly from the inclusive electron \ptt
spectra after subtracting known contributions from photon conversions and
light hadron decays by PHENIX~\cite{phenixcharm}. The spectrum remaining after
subtraction is dominated~\footnote{A possible contribution to the electron
  spectrum from the Drell-Yan process cannot be ruled out at present, though.}
by open charm and beauty contributions. Recently results for an elliptic flow
analysis were shown~\cite{charmv2} of the electrons dominantly from open charm
decays. There is a significant nonzero value in the \ptt range 0.4 - 1.6
GeV/c. This is confirmed by preliminary STAR data~\cite{lamont} that extend
the overall transverse momentum coverage by adding the range \ptt = 1.5 - 3.0
GeV/c. Together, the data paint a consistent picture that indeed the electrons
from open charm decay exhibit elliptic flow, i.e. follow the collective motion
of the light quarks. This would imply that the charm quark thermalizes to a
significant degree. Note that this is a necesarry prerequisite for any
formation of charmed hadrons by statistical hadronization (see below).

In that case also jet quenching should be observed for charmed
hadrons. Indeed, in still preliminary data it was shown recently that electron
spectra, after the subtraction of contributions from conversion and light
hadron decays, show high \ptt suppression for central Au + Au
collisions~\cite{jacak}. The R${\rm _{AA}}$ factor drops practically as low as
for pions at \ptt of 4 GeV/c, i.e. to values of about 0.2. In a recent
publication~\cite{armesto} the suppression for electrons from D meson decay
was studied for different transport coefficients. The preliminary RHIC data
would be consistent with a calculation using a transport coefficient of 14
GeV$^2$/fm (see Fig. 2 of~\cite{armesto}), at the upper end of the range needed
to reproduce R$_{\rm AA}$ for pions. This is very surprizing, in particular also
in view of the fact, that at \ptt of about 4 GeV/c also the contribution of
b-quarks to the electron spectrum should become sizeable.

At top SPS energy, for central Pb + Pb collisions a significant suppression of
J/$\psi$ production was observed in the NA50 experiment~\cite{na501,na50}. This suppression is
compared to the socalled 'normal' nuclear absorption seen also in pA
collisions.  From analysis of all pA data, a cross section for normal nuclear
absorption of 4.1$\pm$0.4 mb was extracted~\cite{na50}. To this normal nuclear
absorption all results from heavy ion collisions can be compared. It turns out
that S + U data as well as data from peripheral Pb + Pb collisions agree with
this normal nuclear absorption curve. For transverse energies above 40 GeV or
a length of nuclear matter seen by the J/$\psi$ of L $\geq$ 7 fm the points
from Pb + Pb collisions fall increasingly below this normal nuclear absorption
curve. Theoretically, the suppression can be explained by disappearance of the
J/$\psi$ (or possibly only the charmonia states that feed it) in a hot colored
medium or by interaction with comovers (chiefly pions), albeit with a very
large density of more than 1/fm$^3$, i.e. a value not deemed achievable for a
hadron gas.

\begin{figure}
\epsfxsize200pt
\figurebox{120pt}{160pt}{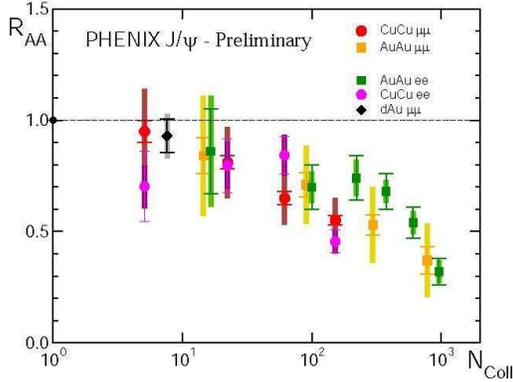}
\caption[]{J/$\psi$ yield in Au + Au and Cu + Cu collisions at top RHIC energy
  normalized to the measured result for pp collisions, scaled with the
  number of inelastic collisions. Results are shown for decays into electron
  and muon pairs at mid- and forward rapidities, respectively. Figure
  from~\cite{akiba}.}
\label{fig:jpsi}
\end{figure}

The first results for J/$\psi$ production in central Au + Au collisions at
RHIC energy came very recently from PHENIX~\cite{akiba} (run4); they are
displayed in Figure~\ref{fig:jpsi}. As compared to pp as well as d + Au
collisions there is a significant suppression. The suppression is, however,
rather similar to the one observed at SPS. This is in contrast to some
predictions~\cite{vogt} that in central Pb + Pb collisions at RHIC only
J/$\psi$ mesons from the corona should survice, i.e. order of 5 \% of the
normal unsuppressed yield. The actually observed yield by far exceeds this
expectation. This could be seen as indication, that at RHIC the energy density
in the QGP is not yet high enough to dissolve the J/$\psi$ but rather only
enough to dissolve higher c\={c} states. Recent results from high temperature
lattice QCD indicate that the J/$\psi$ bound state may only disappear~\cite{karsch2}
in the vicinity of T = 2 T$_c$. 

A maybe more interesting alternative has been proposed~\cite{charmstat}: Even
if the initially formed c\={c} pairs are completely dissociated in the hot
QGP, at hadronization charmed hadrons may form in a statistical fashion by the
same mechanism described above for hadrons involving up, down, and strange
valence quarks. This includes also the formation of charmonia and it was
pointed out in~\cite{charmstat} that the formation probability of J/$\psi$
mesons would grow quadratically with the c\={c} rapidity density. Such
increased reformation of J/$\psi$ by statistical hadronization could possibly
account for a suppression apparently not much stronger than at SPS.

\begin{figure}
\epsfxsize195pt 
\figurebox{120pt}{160pt}{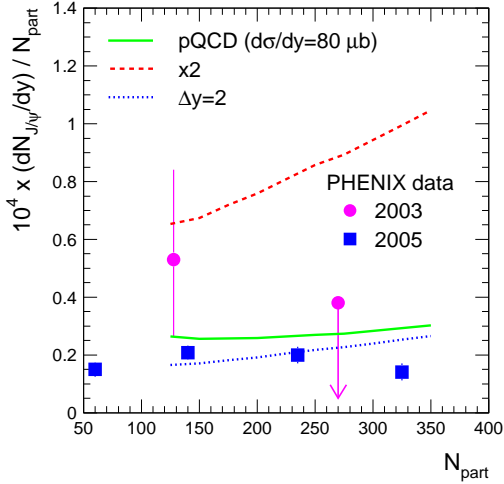}
\caption[]{J/$\psi$ yield per participating nucleon for Au + Au collisions at top RHIC
  energy compared to the yield expected from statistical recombination. Data
  from~\cite{akiba}, statistical hadronization prediction
  from~\cite{charmstat2}. For statistical hadronization a standard interval of
  $\Delta$ y = 1 is used. Calculations are shown for a c\={c} cross section
  per unit rapidity of 80 $\mu$b, of 160 $\mu$b. Also shown: a calculation for
  80 $\mu$b and $\Delta$ y = 2.0.}
\label{fig:jpsistat}
\end{figure}

Figure~\ref{fig:jpsistat} shows the prediction of~\cite{charmstat2} together
with the recent and still preliminary RHIC data for J/$\psi \to$
e$^+$e$^-$. Currently the main uncertainty in this consideration is the
overall charm production yield, which enters quadratically. A calculation is
shown using a cross section per unit rapidity of 80 $\mu$b (400 $\mu$b
integrated) as expected from NLO pQCD~\cite{vogt}. The overall charm
production cross section at RHIC energy has so far been measured indirectly by
PHENIX~\cite{phenixcharm} for $\sqrt{\rm s}$ = 130 and 200 GeV from the
inclusive electron spectra in the way described above for Au + Au collisions
and, at full RHIC energy, also for d + Au and pp
collisions~\cite{phenixcharm2}.  It is found that the integrated charm cross
section, when scaled with the number of binary collisions, agrees for all
three collision systems. The value is about 50\% above the NLO pQCD
calculation~\cite{vogt} but agrees within errors. On the other hand, in STAR,
D mesons have been reconstructed via their hadronic decay to K$\pi$ in d + Au
collisions and a charm cross section per nucleon nucleon collision has been
extracted~\cite{starcharm}. It is twice as large as the PHENIX value by nearly
two standard deviations. The experimental situation concerning open charm
production needs to be improved before the J/$\psi$ puzzle can be better
addressed. Only measurements at LHC will unambiguously clarify the role of
statistical hadronisation of charm, since with this mechanism a significant
J/$\psi$ enhancement in Pb + Pb collisions at LHC was
predicted~\cite{charmstat2} instead of suppression. Such an enhancement would
be an unambiguous signal of deconfinement.

\section{Summary and Outlook}

Hadron yields are found to be in chemical equilibrium. For top SPS energy and
up this can be achieved by multi-particle collisions in the direct
vicinity of T$_c$ and hence the observed chemical equilibration temperature is
an experimental measure of the critical temperature for the phase transition.

At RHIC energies, spectra and azimuthal correlations are quantitatively
described by hydrodynamics. This requires rapid local thermalization and high
initial energy densities more than tenfold above the calculated critical
energy density for the phase transition between hadronic matter and QGP.

High \ptt hadrons are suppressed in central Au + Au collisions and this is a
medium effect. Jet quenching in a hot color charged medium was predicted,
modelling of the data with high parton density is successful. There are some
indications of valence quark coalescence in hadronic observables.

The observations that lead to the CERN press release are confirmed by the 
RHIC experiments. Beyond this additional features are observed that start to
probe the properties of the new state of matter. Much progress in this
direction is expected from the high luminosity RHIC data just starting to
appear and, from 2007, from the LHC heavy ion program.

\section*{Acknowledgements}
I thank Peter Braun-Munzinger for numerous enlightening discussion.

\end{document}